\documentclass[twoside,10pt]{article}
\usepackage{epsfig}
\usepackage{lineno}
\usepackage{cite}
\usepackage{footnote}
\usepackage{float}

\newcommand{\be}{\begin{equation}}
\newcommand{\ee}{\end{equation}}
\newcommand{\bea}{\begin{eqnarray}}
\newcommand{\eea}{\end{eqnarray}}

\begin{document}

\title{ \vspace{1cm} Isotopic $^{32}$S/$^{33}$S ratio as a diagnostic of presolar grains from novae}
\author{A. Parikh$^{1,2,3,4,*}$, K. Wimmer$^{5,3,4}$, T. Faestermann$^{3,4}$, R. Hertenberger$^{4,6}$, J. Jos\'e$^{1,2}$, H.-F. Wirth$^{4,6}$,\\ C. Hinke$^{3,4}$, R. Kr\"ucken$^{7,3,4}$, D. Seiler$^{3,4}$, K. Steiger$^{3,4}$, K. Straub$^{3,4}$\\
\\
\small $^1$Departament de F\'isica i Enginyeria Nuclear, \\ \small Universitat
Polit\`ecnica de Catalunya, EUETIB, E-08036 Barcelona, Spain\\
\small $^2$Institut d'Estudis Espacials de Catalunya (IEEC), E-08034 Barcelona, Spain\\
\small $^3$Physik Department E12, Technische Universit\"at M\"unchen, D-85748 Garching, Germany\\
\small $^4$Maier-Leibnitz-Laboratorium der M\"unchner Universit\"aten (MLL), D-85748 Garching, Germany\\
\small $^5$Department of Physics, Central Michigan University, Mount Pleasant, Michigan 48859, USA\\
\small $^6$Fakult\"at f\"ur Physik, Ludwig-Maximilians-Universit\"at M\"unchen, D-85748 Garching, Germany\\
\small $^7$TRIUMF, Vancouver, BC, Canada V6T 2A3}
\maketitle

\begin{abstract} 

Measurements of sulphur isotopes in presolar grains can help to identify the astrophysical sites in which these grains were formed.  A more precise thermonuclear rate of the $^{33}$S($p,\gamma$)$^{34}$Cl reaction is required, however, to assess the diagnostic ability of sulphur isotopic ratios.   We have studied the $^{33}$S($^{3}$He,d)$^{34}$Cl proton-transfer reaction at 25 MeV using a high-resolution quadrupole-dipole-dipole-dipole magnetic spectrograph.  Deuteron spectra were measured at ten scattering angles between 10$^{\circ}$ and 55$^{\circ}$.  Twenty-four levels in $^{34}$Cl over E$_{x}$=$4.6 - 5.9$ MeV were observed, including three levels for the first time.  Proton spectroscopic factors were extracted for the first time for levels above the $^{33}$S+p threshold, spanning the energy range required for calculations of the thermonuclear $^{33}$S(p,$\gamma$)$^{34}$Cl rate in classical nova explosions.  We have determined a new $^{33}$S($p,\gamma$)$^{34}$Cl rate using a Monte Carlo method and have performed new hydrodynamic nova simulations to determine the impact on nova nucleosynthesis of remaining nuclear physics uncertainties in the reaction rate.  We find that these uncertainties lead to a factor of $\leq5$ variation in the $^{33}$S(p,$\gamma$)$^{34}$Cl rate over typical nova peak temperatures, and variation in the ejected nova yields of S--Ca isotopes by $\leq20\%$.  In particular, the predicted $^{32}$S/$^{33}$S ratio is $110-130$ for the nova model considered, compared to $110-440$ with previous rate uncertainties.  As recent type II supernova models predict ratios of $130-200$, the $^{32}$S/$^{33}$S ratio may be used to distinguish between grains of nova and supernova origin.

\end{abstract}


\vspace{15 mm}
{\footnotesize *Corresponding author.  Email address: anuj.r.parikh@upc.edu (A. Parikh). }

\newpage

\twocolumn

\section{Introduction}
\label{introduction}

Classical nova explosions occur through thermonuclear ignition in a shell of
hydrogen-rich material accreted by a white dwarf star in a binary star system (for reviews, see e.g., Refs.\cite{Bod08, Jos07a, Par14}).  Several hundred Galactic novae
have been discovered to date, with roughly five events discovered per year. Light curves for these
events peak at $\approx10^{4} - 10^{5}$ times the solar luminosity and persist for intervals of
$\approx$days to several months.  A nova explosion will typically eject $\approx10^{-4} - 10^{-5}$ solar masses of material into the interstellar
medium.  

Nucleosynthesis predictions of current models are in broad agreement with the observed (elemental) composition of nova ejecta.  Models find that nova explosions only on very massive oxygen-neon white
dwarfs, reaching peak temperatures $T\geq 0.3$ GK, are likely to synthesize the heaviest 
species observed (i.e., in the Si -- Ca mass range\cite{Sta98,Jos01,Sta09}).  More precise abundance predictions in this mass range could be obtained through improvements in a limited number of nuclear reaction rates, including $^{30}$P($p,\gamma$)$^{31}$S and $^{33}$S($p,\gamma$)$^{34}$Cl\cite{Jos01,Ili02,Par11,Fal13, Dow13, Kel13}.

More stringent tests of nucleosynthesis predictions could be made through the use of isotopic, as opposed to elemental, constraints on the composition of nova ejecta.  Such information could be provided through the detection of $\gamma$-rays from the decay of radioisotopes produced during the explosion, or through measurements of presolar grains -- microscopic grains embedded within primitive meteorites.  Unfortunately, in the former case, only upper limits on nuclear $\gamma$-ray emission from novae have been obtained to date (all of which are fully compatible with theoretical predictions); and in the latter case, only a handful of measured grains exhibit signs of nova nucleosynthesis\cite{Ama01, Ama02, Jos04} -- and even these may also be consistent with origin in type II supernovae\cite{Nit05,Jos07b}.  Measurements of sulphur isotopes in grains (see, e.g., Refs.\cite{Hop10,Gyn12,Hop12} for recent progress) could provide a valuable means of discriminating between grains from novae and other stellar environments when used in concert with other isotopic ratios indicative of nova nucleosynthesis (such as $^{12}$C/$^{13}$C, $^{14}$N/$^{15}$N and $^{29,30}$Si/$^{28}$Si\cite{Jos04}).  In particular, current type II supernova models predict ejected material with a $^{32}$S/$^{33}$S ratio (in terms of mass fractions, here and throughout) of $\approx130-200$\cite{Chi13}, while recent nova models predict ejecta with a ratio of $\approx110-440$, depending on the $^{33}$S($p,\gamma$)$^{34}$Cl rate used\cite{Fal13}.  We note that the precision of measurements of these ratios in grains varies from $\approx1\%$ (for carbon, nitrogen and silicon isotopes\cite{Nit05}) to $\approx10\%$ for sulphur isotopes\cite{Hop12}.    As such, a more precise $^{33}$S($p,\gamma$)$^{34}$Cl rate is clearly needed to better constrain predictions of sulphur abundances in novae, so as to assess the utility of the $^{32}$S/$^{33}$S ratio as a site discriminant.

At temperatures encountered within nova explosions, the thermonuclear rate of the
$^{33}$S($p,\gamma$)$^{34}$Cl reaction is dominated by contributions from narrow resonances
within $\approx600$ keV of the $^{33}$S+p threshold in $^{34}$Cl (S$_{p}$($^{34}$Cl) = 5143.2 keV\cite{AME12}). Direct measurements of this reaction\cite{Das77,Waa83,Fre11,Fal13} have determined resonance energies (which enter exponentially in the rate), ($p,\gamma$) resonance strengths (which enter linearly in the rate), spin-parity values and $\gamma$-decay schemes for levels with E$_{x}$($^{34}$Cl) $> 5.43$ MeV, with the reaction cross-section limiting the lowest energies at which measurements have been performed.  Indirect studies have determined the energies of 11 levels in $^{34}$Cl between the proton-threshold and $5.43$ MeV\cite{End90, Par09,NDS}, but no experimental information exists for the corresponding resonance strengths.  Parameterizations and statistical model calculations have also been used to determine this rate in the past\cite{Jos01,Ili01,Rau00}.

Strengths of the low-energy resonances below 5.43 MeV are needed for an improved calculation of the $^{33}$S($p,\gamma$)$^{34}$Cl rate.  The ($p,\gamma$) strength of a resonance is proportional to its proton partial width.  The proton partial width can be determined using a proton-transfer spectroscopic factor extracted from angular distribution measurements of the $^{33}$S($^{3}$He,d) reaction.  The $^{33}$S($^{3}$He,d)$^{34}$Cl reaction has been studied before, but only for states with E$_{x}$($^{34}$Cl) $< 4.64$ MeV\cite{erskine71}.  In the present work, we have measured the $^{33}$S($^{3}$He,d)$^{34}$Cl reaction to provide, for the first time, proton spectroscopic factors for states above the proton threshold in $^{34}$Cl.  With these, we have determined the $^{33}$S($p,\gamma$) resonance strengths of the key low-energy resonances.  We have then used a Monte Carlo method to calculate a new thermonuclear $^{33}$S($p,\gamma$)$^{34}$Cl rate with statistically meaningful uncertainties.  Finally, we have computed hydrodynamic nova models to test the impact of remaining nuclear physics uncertainties in this rate on nova nucleosynthesis predictions.  In particular, we evaluate whether or not measurements of the $^{32}$S/$^{33}$S ratio in grains would provide an effective means of identifying grains of nova paternity.

\section{Experiment}
\label{experiment}

The $^{33}$S($^{3}$He,d)$^{34}$Cl reaction was measured at the Maier-Leibnitz-Laboratorium in Garching, Germany.  A 25-MeV $^{3}$He$^{2+}$ beam ($I \approx 300 - 600$ nA) was produced using an ECR-like ion source \cite{ECR} and an MP tandem accelerator.  This beam was brought to the target position of a quadrupole-dipole-dipole-dipole magnetic spectrograph ($ \Delta E/E \approx 2 \times 10^{-4}$ \cite{Q3D}).  Targets were prepared at the Technische Universit\"at M\"unchen and included Ag$_{2}$$^{33}$S (20 $\mu$g/cm$^{2}$, enriched to 99.9\% $^{33}$S) deposited upon a foil of $^{12}$C; Ag$_{2}$$^{nat}$S (20 $\mu$g/cm$^{2}$, natural sulphur) deposited upon a foil of $^{12}$C; a 50 $\mu$g/cm$^{2}$ Ag foil; a 7 $\mu$g/cm$^{2}$ $^{12}$C foil; and $^{24}$Mg (20 $\mu$g/cm$^{2}$, enriched to 99.92\% $^{24}$Mg) deposited upon a foil of $^{12}$C.  All $^{12}$C foils were enriched to 99.99\% and of thickness 7 $\mu$g/cm$^{2}$.  Light reaction products entered the spectrograph through a rectangular aperture of 7.0 msr, were analyzed according to their momenta, and were focused onto a multiwire gas-filled proportional counter backed by a plastic scintillator \cite{FPD}.  Deuterons were identified and selected through energy loss and residual energy information from the focal-plane detection system, and deuteron spectra of focal-plane positions were then produced for further analysis.  Measurements were made at spectrograph angles of 10$^\circ$, 15$^\circ$, 20$^\circ$, 25$^\circ$, 30$^\circ$, 35$^\circ$, 40$^\circ$, 45$^\circ$, 50$^\circ$ and 55$^\circ$.  The beam current was integrated using a Faraday cup placed at 0$^\circ$ in the target chamber.

\section{Data and Analysis}
\label{data}

\begin{figure}[H]
\begin{center}
\includegraphics[scale=0.45]{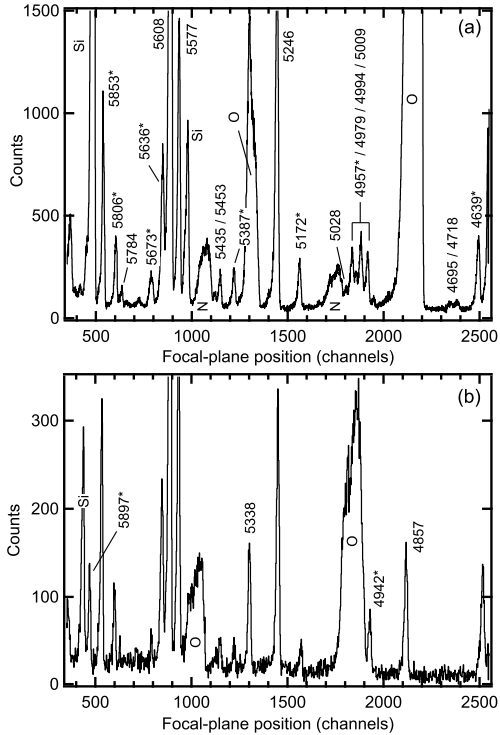}
\caption{Focal-plane position spectra of deuterons from the $^{33}$S($^{3}$He,d)$^{34}$Cl reaction at 25 MeV and (a) $\theta_{lab}$ = 15$^\circ$ and (b) $\theta_{lab}$ = 25$^\circ$.  Peaks are labeled by the corresponding $^{34}$Cl excitation energies in keV; states used in the energy calibration of the focal plane are denoted with an asterisk.  Contaminant groups from ($^{3}$He,d) reactions on $^{14}$N, $^{16}$O, and $^{28}$Si present in the target are indicated.}
\label{fig1}
\end{center}
\end{figure}

Figure \ref{fig1} shows deuteron position spectra measured with the Ag$_{2}$$^{33}$S
target at spectrograph angles of 15$^{\circ}$ and 25$^{\circ}$. Contaminant
groups due to ($^{3}$He,d) reactions on $^{14}$N, $^{16}$O and $^{28}$Si present in
the target are evident, and these were identified and
characterized through both kinematic analysis at the measured
angles and measurements with the $^{12}$C, Ag and Ag$_{2}$$^{nat}$S targets.  These
spectra were analyzed using least-squares fits of multiple
Gaussian or exponentially modified Gaussian functions with
a constant or linearly-varying background. Consistent excitation energies were
determined using each of these prescriptions.   The energy resolution
was determined to be $\approx 12$ keV full width at
half maximum from the widths of fits to isolated peaks in these spectra.

The focal plane was initially calibrated at each measured angle
using well-resolved, known states in $^{25}$Al populated through the ($^{3}$He,d) reaction on the $^{24}$Mg target.  These low-energy states (1.6 $<$ E$_{x}$($^{25}$Al) $<$ 3.1 MeV) are known to a precision of better than 1 keV\cite{Fir09}.  Second-degree polynomial
least-squares fits of deuteron radius-of-curvature to focal-plane
position were obtained at each angle using this information,
and these fits were then used to determine excitation energies
for states in $^{34}$Cl populated via the $^{33}$S($^{3}$He,d) reaction.  Clearly resolved, strongly populated states in $^{34}$Cl with energies known to $\leq$1.5 keV\cite{NDS,End90} (and in agreement with those energies determined through the $^{25}$Al calibration) were then used as part of an internal calibration at each angle to finally determine the energies of states in $^{34}$Cl.  Table \ref{table1} lists  excitation energies determined in the present work, along with
uncertainties due to counting statistics, reproducibility
among angles and uncertainties in the energies of the calibration states.  The $^{34}$Cl states used with the internal calibrations are also indicated in Table \ref{table1}.  Each energy from the present work is averaged
over values from at least three angles (depending upon, e.g., the presence of contaminant states) unless otherwise indicated in the table.

\begin{figure}[H]
\begin{center}
\includegraphics[scale=0.35]{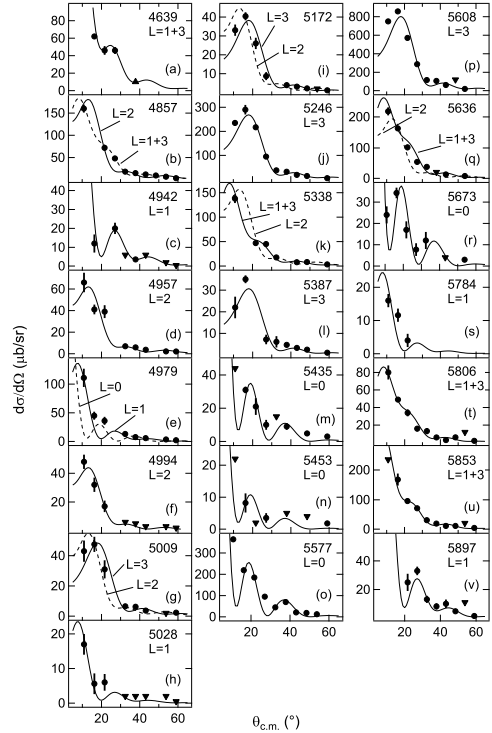}
\caption{ Deuteron angular distributions measured with the
$^{33}$S($^{3}$He,d)$^{34}$Cl reaction at 25 MeV.  Upper and lower limits are indicated using triangles.  Panels (a--v) are labeled with $^{34}$Cl excitation energies
(in keV) and the transferred orbital angular
momentum $L$ of calculated differential cross sections fit to the data.}
\label{fig2}
\end{center}
\end{figure}

Figure \ref{fig2} shows measured angular distributions for states in $^{34}$Cl populated through the $^{33}$S($^{3}$He,d)
reaction.  Proton-transfer spectroscopic factors $C^{2}S$ have been extracted for each state by fitting these differential cross sections using theoretical calculations from the code FRESCO~\cite{thompson88} for different values of the transferred orbital angular momentum $L$ (including mixed transitions).  Constraints implied by previous measurements of resonance strengths or spin-parity values, when available, were considered when determining the adopted theoretical calculation.  For several states $C^{2}S$ values for two equally plausible transitions were found.  The results are summarized in Table \ref{table1}; a systematic uncertainty of $\approx10\%$ due to uncertainty in the target thickness should be included separately.  Different sets of optical model parameters, scaled to the present target mass and beam energy, were adopted in the calculations to assess additional associated systematic uncertainties in the spectroscopic factors (see Section \ref{thermonuclearrate} as well).  These included global parameters from Ref.~\cite{perey76} (for both the deuteron and the $^3$He channels); parameters given in Ref.~\cite{erskine71} (who also measured the ($^{3}$He,d) reaction, albeit at a lower beam energy of 14 MeV); and parameters from Ref.~\cite{pang09} and Ref.~\cite{daehnick80} for the $^3$He entrance channel and the deuteron exit channel, respectively.  Values of the extracted spectroscopic factors were found to be relatively insensitive to these different sets of optical model parameters, with variations of only $\approx5-10\%$ depending on the state.

\section{Discussion}
\label{discussion}

\subsection{Spectroscopy of $^{34}$Cl}
Excitation energies for $^{34}$Cl levels observed in the present work are in excellent agreement with values previously reported in the literature (see Table \ref{table1}).  We have substantially reduced the uncertainties in the energies of four levels at 4979, 5009, 5028, and 5608 keV and confirmed two very recently observed levels\cite{Fal13} at 5435 and 5453 keV.    In addition we have measured three new levels at 4857, 5246 and 5338 keV.  The observation of new levels in this energy range is not unexpected given that levels from neither the $^{33}$S($^{3}$He,d) nor the $^{33}$S($p,\gamma$) reaction have been previously reported for E$_{x}$ $>$ 4.63 MeV\cite{erskine71} and E$_{x}$ $<$ 5.43 MeV\cite{Das77,Waa83,Fal13}, respectively.  The selectivity of the $^{33}$S($^{3}$He,d)$^{34}$Cl reaction is evident from Table \ref{table1}.  For example, of the 37 states previously identified in the range $4.9 < E_{x} < 5.9$ MeV, only 18 were observed in the present work.  In particular, for energies where direct measurements exist (E$_x > 5.43$ MeV), we populate levels previously observed through the $^{33}$S($p,\gamma$) reaction almost exclusively (see Table \ref{table2}).  The only exceptions are the states at 5607 and 5763 keV, where the former is observed here but not reported in the ($p,\gamma$) study of Ref.\cite{Waa83}, and the latter is not observed here but is reported in Ref.\cite{Waa83}.  These exceptions are likely due to the low associated ($p,\gamma$) resonance strengths of these two states (see Section \ref{thermonuclearrate}).      

Spin-parity constraints from the present work, based upon the adopted orbital angular momentum transfer $L$ for each state, are in general consistent with previous assignments.  Only tentative assignments exist for two of the three exceptions, at E$_{x}$ = 5578 and 5806 keV\cite{NDS}.  The third exception, at 4942 keV, is a known $1^{+}$ state; we prefer a transition with $L=1$, leading to an assignment with negative parity for this state.  The discrepancy is perhaps due to the very limited angular distribution measured here, especially at low angles (see Fig. \ref{fig2}c), because of the presence of states from target contaminants. 


\subsection{Thermonuclear rate}
\label{thermonuclearrate}

Under the assumption of narrow, isolated resonances, the resonant rate $\langle \sigma v\rangle$ of the $^{33}$S($p,\gamma$)$^{34}$Cl reaction can be calculated (in cm$^{3}$ s$^{-1}$ mol$^{-1}$) as\cite{Ili07}
\begin{equation}
\label{eq_rate}
\footnotesize{
N_{A}\langle \sigma v\rangle=1.5399\times10^{11}(\mu T_{9})^{-3/2} \sum\limits_{i} (\omega\gamma)_{i}\exp(-11.605 E_{R,i}/T_{9})
}
\end{equation}
where $N_{A}$ is the Avogadro constant, $T_{9}$ is the temperature in GK, $\mu$ is the reduced mass of the $^{33}$S+p system in u, and the E$_{R,i}$ are resonance energies in MeV.  The resonance strength $\omega\gamma$ (in MeV in Eq. \ref{eq_rate}) can be expressed as
\begin{equation}
\label{eq_wg}
\omega\gamma  = \frac{2J_{R}+1}{(2J_{p}+1)(2J_{t}+1)} \cdot \frac{\Gamma_p\Gamma_\gamma}{\Gamma_{tot}}
\end{equation}
where $J_{R}$, $J_{p} =1/2$, and $J_{t} =3/2$ are the spins of the resonance in $^{34}$Cl, the proton, and the ground state of $^{33}$S, respectively. The total width $\Gamma_{tot}$ of a resonance is the sum of the proton and $\gamma$-ray partial widths ($\Gamma_{p}$ and $\Gamma_{\gamma}$, respectively) for proton-threshold states. The sum in Eq. \ref {eq_rate} allows for the contributions of all resonant states through which the reaction may proceed at the chosen temperature.  

We have determined a new thermonuclear rate for the $^{33}$S($p,\gamma$)$^{34}$Cl reaction using experimental results from the present $^{33}$S($^{3}$He,d) measurement and previous direct $^{33}$S($p,\gamma$) studies.  For $E_{R} = 790 - 1940$ keV, we have used the energies, strengths and associated uncertainties of the 49 $^{33}$S($p,\gamma$) resonances in Ref.\cite{End90}.  For $E_{R} = 290 - 790$ keV we have used strengths from previous direct ($p,\gamma$) measurements as listed in Table \ref{table2}.  For $E_{R} < 290$ keV we have determined resonance strengths using our measured spectroscopic factors.  First,  proton partial widths were calculated according to $\Gamma_p = C^2S \cdot \Gamma^{sp}_p$,
where $\Gamma^{sp}_p$ is the single-particle width.  Values for $\Gamma^{sp}_p$ were determined with both the program WSPOT\cite{WSPOT}, using a Woods-Saxon potential with a radius of $R=4.08$~fm and a diffuseness of $a=0.6$~fm,
and through the expression\cite{Ili07}
\[ \Gamma^{sp}_{p} = \frac{2\hbar^2}{\mu r^2}P_{L}\theta_{sp}^2  \]
where $r$ is the interaction radius, $P_{L}$ is the
penetrability of the Coulomb and centrifugal barriers for orbital
angular momentum transfer $L$, and $\theta_{sp}^2$ is the
single-particle reduced width\cite{Ili97}.  The two methods gave $\Gamma^{sp}_{p}$ values that agreed to better than 10\%.  Proton partial widths found using the above prescription are listed in Table \ref{table2}.  Strengths for the four resonances with $E_{R} < 290$ keV (and the resonance at 464 keV) were then determined using Eq. \ref{eq_wg} with the reasonable assumption that $\Gamma_{p} << \Gamma_{\gamma}$ for these proton-threshold states.  Uncertainties for these five strengths were found using low and high compatible $J$ values (see Table \ref{table2}), the uncertainties in the $C^{2}S$ values (see Table \ref{table1} and Section \ref{data}), and, for the 28 and 195 keV resonances, $\omega\gamma$ values arising from both possible transitions.  Furthermore, we adopted additional, conservative, systematic uncertainties of a factor of four (enhancement) and a factor of two (reduction) for the highest and lowest possible strengths, respectively, calculated for each of these five resonances.  As a result we considered, for example, strengths of $1.7\times10^{-6} - 1.5\times10^{-4}$ meV and $9.4\times10^{-7} -1.2\times10^{-5}$ meV, for the 195 and 244 keV resonances, respectively.  This systematic uncertainty accounts for discrepancies when comparing strengths determined using the above treatment to strengths measured in previous direct studies\cite{Waa83,Fal13}, for $E_{R}>290$ keV.  Measurements of $\Gamma_{\gamma}$ for these states (assumed here to be $>> \Gamma_{p}$) would help to clarify the source of these discrepancies.  The adopted systematic uncertainties are also suggested from comparing spectroscopic factors for the single common state observed in both the present and previous measurements of the $^{33}$S($^{3}$He,d) reaction.  Erskine et al.\cite{erskine71} measured $^{34}$Cl states with E$_{x}< 4.64$ MeV, finding $C^{2}S=0.05$, $L=1$ for the $J=2^{-}$\cite{NDS}, 4639 keV state.  While our data suggest a $L=1+3$ mixed transition for this state, a fit using a pure $L=1$ transition gives $C^{2}S=0.019$.  

\begin{figure}[H]
\centering
\includegraphics[scale=0.4]{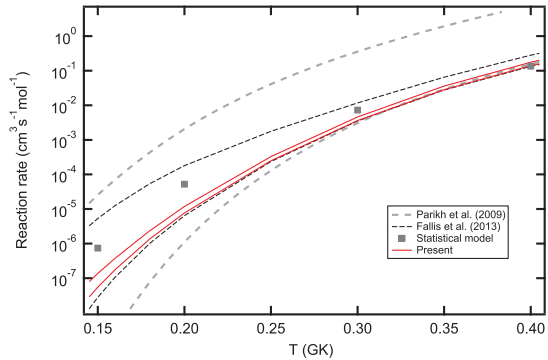}
\caption{(Color online)  Thermonuclear $^{33}$S($p,\gamma$)$^{34}$Cl rates over typical nova peak
temperatures.  Low and high rates from the present work (solid red lines), rates A and D from Fallis et al.\cite{Fal13} (black dashed lines), upper and lower rates estimated in Parikh et al.\cite{Par09} (grey dashed lines) and a rate calculated using a statistical model\cite{Rau00,Ili01,Fal13} (filled grey squares) are all indicated. }
\label{fig_rate}
\end{figure}

A Monte Carlo method, as implemented through the STARLIB library\cite{STARLIB}, was used to determine low, median, and high thermonuclear $^{33}$S($p,\gamma$)$^{34}$Cl rates using these resonance parameters and the associated uncertainties in $E_{R}$ and $\omega\gamma$.  
The systematic uncertainties in $\omega\gamma$ discussed above were treated as random and independent for the relevant levels.  
Hereafter ``low rate" and ``high rate" refer to the 0.16 and 0.84 quantiles of the cumulative rate distributions at each temperature $T$, as discussed in Ref.\cite{Lon10}.  The direct-capture component of
the rate was estimated with $S_{DC}(0) = 30$ keV-b\cite{Ili01,Ili07,Par09} and is negligible above $\approx0.03$ GK, at which it is less than $\approx 1 $\% of the low rate.   The low and high rates at the relevant temperatures are dominated by contributions from the resonances at 292, 310, 434, 493 and 530 keV.  For example, at 0.3 GK, even if the maximum strengths are assumed for all of the resonances with $E_{R} < 290$ keV and 464 keV (see above), the \emph{sum} of the contributions of these five resonances is only 4\% of the contribution of the 434 keV resonance to the rate.  As such, our results below do not hinge upon our estimated systematic uncertainties for the strengths of the five resonances with $E_{R} < 290$ keV and 464 keV.  

In Fig.~\ref{fig_rate}, our low and high rates are compared to previous rates, including that from a statistical model calculation\cite{Rau00,Ili01} normalized to an experimental rate at 2 GK\cite{Fal13}.  At 0.1, 0.3 and 1 GK the ratio between the present high and low rates is 5.3, 1.3, and 1.3, respectively.  The rate uncertainty has been reduced significantly relative to that presented in previous studies\cite{Par09,Fal13}.  For example, the ratio between rates A and D of Fallis et al.\cite{Fal13} (see Fig. \ref{fig_rate}) at 0.1 and 0.3 GK is 16000 and 3.3, respectively.  We note that since $^{34}$Cl has a metastable state (E$_{x} = 146$ keV, t$_{1/2} = 32$ min), one should, in principle, distinguish between proton-capture rates to the ground and metastable states.  Due to the lack of sufficient information
on the relative population of these states following the $\gamma$-decay of resonances with E$_{x} < 5.58$ MeV (i.e., relevant for the reaction rate at nova temperatures, of interest in this work), we have only determined the total $^{33}$S($p,\gamma$)$^{34}$Cl rate.

\subsection{Astrophysical Impact}

In order to assess the impact on nova nucleosynthesis predictions of remaining uncertainties in the $^{33}$S($p,\gamma$)$^{34}$Cl rate, we have performed simulations with SHIVA \cite{JH98}, a one-dimensional
(spherically symmetric), hydrodynamic, implicit, Lagrangian code used extensively in the modeling of stellar explosions such
as classical novae and type I X-ray bursts. Two different hydrodynamic models of nova outbursts have been computed, from the early 
accretion stage through the explosion, expansion and ejection stages. These models were identical except for the $^{33}$S($p,\gamma$)$^{34}$Cl rate adopted: one model used the ``low rate" from the present work, and the other used the ``high rate" (see Fig. \ref{fig_rate}).  In both cases a 1.35 M$_{\odot}$ oxygen-neon white dwarf has been assumed, and the solar-like accreted material has been pre-mixed with material from the outer layers of the white dwarf at a level of 50\%, to mimic 
mixing at the core-envelope interface \cite{Cas11b}. A characteristic initial luminosity of $10^{-2}$ L$_\odot$ and
a mass-accretion rate of $2 \times 10^{-10}$ M$_\odot$ yr$^{-1}$ were also assumed.  As shown in Table \ref{table3}, differences in the nucleosynthesis of the two models appear in the S--Ca mass region, although these are decidedly modest (e.g., $20\%$ for $^{33}$S).  Significantly larger differences for $^{33}$S were observed when comparing the nucleosynthesis from nova models using previous $^{33}$S($p,\gamma$) rate uncertainties (e.g., a factor of 4.2 for rates A and D in Fallis et al.\cite{Fal13}).   Since $^{34}$Cl was included as a single species in the reaction network, we note that yields of nuclei with $A \geq 34$ may differ somewhat from those in Table \ref{table3} if separate production and destruction rates for the ground and metastable states of $^{34}$Cl were included; however, reliable estimates of these rates are not currently available.  As such, we now focus on the $^{32}$S/$^{33}$S isotopic ratio.

The present uncertainty in the $^{33}$S($p,\gamma$) rate gives a range of $110-130$ for the predicted $^{32}$S/$^{33}$S ratio from our nova models.  This tight constraint allows us to examine the utility of using this observable to identify meteoritic grains originating from classical nova explosions, and in particular, to differentiate between grains formed following type II supernova and nova explosions.  Primary characteristics used previously to help classify nova grains have included ratios of, among others, $^{12}$C/$^{13}$C, $^{14}$N/$^{15}$N, $^{29}$Si/$^{28}$Si and $^{30}$Si/$^{28}$Si; nova models predict ranges of roughly $0.3-3$, $0.1-10$, $0.01 - 0.15$, $0.01 - 0.15$, and $0.007 - 0.6$, respectively\cite{JH98, Jos04, Nit05}.  Recent type II supernova models, on the other hand, predict ranges of $200-3000$, $1000-14000$, $0.02-0.03$, and $0.02 - 0.04$ for these ratios\cite{Chi13}.  Low measured $^{12}$C/$^{13}$C and $^{14}$N/$^{15}$N ratios should ostensibly link grains to nova explosions\cite{Ama01}; however, the reliability of these signatures has been debated\cite{Nit05,Jos07b}.  Additional signatures are evidently essential.   Our nova model predicts a $^{32}$S/$^{33}$S ratio of $110-130$, which may be compared to predictions of $130-200$ from recent type II supernova models for progenitors of $13 - 120$ M$_{\odot}$\cite{Chi13}.  Hence, the $^{32}$S/$^{33}$S ratio is a signature that can help to distinguish between grains from classical novae and type II supernovae.  The present reduction in the uncertainty of the $^{33}$S($p,\gamma$) thermonuclear rate is key to this result as the previous rate uncertainty gave a range of $^{32}$S/$^{33}$S ratios from novae ($110-440$\cite{Fal13}) that encompasses the predictions from supernova models.   

\section{Conclusions} 
\label{conclusions}
      
To better constrain the thermonuclear rate of the $^{33}$S($p,\gamma$)$^{34}$Cl reaction, we have measured the $^{33}$S($^{3}$He,d)$^{34}$Cl proton-transfer reaction at 25 MeV using a high-resolution magnetic spectrograph.  We have observed three new levels in $^{34}$Cl and, for the first time, extracted proton spectroscopic factors for all levels in $^{34}$Cl relevant for calculations of the $^{33}$S($p,\gamma$)$^{34}$Cl rate over temperatures involved in classical nova explosions.  Using these results we have reduced the uncertainty in this reaction rate considerably, from four orders of magnitude at 0.1 GK to factors of 5.3 and 1.3, at temperatures of 0.1 and 1 GK, respectively.   Hydrodynamic nova models performed demonstrate that the present uncertainty in the $^{33}$S($p,\gamma$)$^{34}$Cl rate results in uncertainties of $\leq20\%$ for nova yields of species between S and Ca.  In particular, the adopted nova model predicts a $^{32}$S/$^{33}$S ratio of $110-130$ in the ejecta, as opposed to ratios of $110-440$ using previous rate uncertainties.  Given that recent type II supernova models predict a ratio of $130-200$, the $^{32}$S/$^{33}$S ratio may be used to distinguish between presolar grains of nova and supernova origin when used in conjunction with other predicted isotopic signatures.

\section*{Acknowledgments}

It is a pleasure to thank the crew of the MLL tandem accelerator.  This work was supported by the DFG Cluster of Excellence ``Origin and Structure of the Universe" (www.universe-cluster.de).

\begin{table}[ht]
\centering
  \caption{Nuclear structure of $^{34}$Cl for states within $4.6 < E_{x} < 5.9$ MeV.  The first two columns give weighted averages of $E_{x}$ (in keV) and $J^{\pi}$ assignments from previous work.  Resonance energies from Ref.\cite{Fal13} have been converted to $E_{x}$ using atomic masses and S$_{p}$($^{34}$Cl)$=5143.2$ keV\cite{AME12}.   The final three columns give $E_{x}$ and extracted proton spectroscopic factors $C^{2}S$ for the adopted orbital angular momentum transfer values $L$, as determined in the present work.} 
\vspace{4 mm}
\resizebox{0.7\textwidth}{!}{\begin{minipage}{\textwidth}

\begin{tabular}{ |ll|lll| }
  \hline
  \multicolumn{2}{|c|}{Previous work}&\multicolumn{3}{|c|}{Present work} \\
  E$_{x}$\cite{NDS,Fal13}&J$^{\pi}$\cite{NDS}&E$_{x}$& L & C$^{2}$S \\
  \hline
  4638.9(4) & $2^{-}$   & *  & 1+3 &0.010(2)+0.017(3)\\
  4670(40) & ($3^{-}$)   &   &  & \\
  4680(10) &  &    &  &   \\
  4695.7(2) & $0^{+},1,2,3,4^{+}$   & 4695(2)$^{a}$   &  & \\
  4717.4(6) & $1^{+},2$   & 4718(2)$^{a}$   &  & \\
  4743.15(11) & $6^{-}$   &  &  &   \\
  (4788(7)) & 3 to $7^{+}$   &  &   &  \\
  4824.2(1) & $5^{+}$   &  &    & \\
  &    & 4857(2)  & 2 &0.230(8) \\
  &    &  & 1+3 &0.0059(5)+0.034(3) \\
  4927(4) & 0 to $4^{+}$   &  &    & \\
  4941.9(4) & $1^{+}$   & *  & 1 &0.0060(7) \\
  4957.3(11) & $1^{+},2^{+}$   & * & 2 &0.079(5) \\
  (4970(40)) & $(0)^{+}$   &  &  &   \\
  4971(11) & $1^{+},2^{+}$   & 4979(2)  & 1 &0.0058(4) \\
   &   &   & 0 &0.0095(7) \\
  4995.6(3) & $1^{+},2^{+}$   & 4994(2)  & 2&0.056(5)  \\
  5010(13) & (0 to 4)$^{+}$   & 5009(2) & 2&0.070(3)  \\
  &    & & 3&0.020(1)  \\
  (5020) &    & 5028(2) & 1 &0.0010(1) \\
  5061(4)  &  &  &  &   \\
  5093(4)   &  &  &  &   \\
  5154(4)   &  &  &  &  \\
  5171.6(3) & 4   & *  & 3 &0.0159(7) \\
  &   &  & 2 &0.057(3) \\
  5233(4) &  &    &  &   \\
  &    & 5246(2)  & 3 &0.112(3) \\
  5263(4)   &  &  &  &   \\
  5278(7) & $5^{+},6^{+},7^{+}$     &  &  & \\
  5292 &  &  &  &     \\
  5314.93(17) & $7^{+}$     &  &  & \\
  5326(4) &  &  &  &     \\
  &    & 5338(2) & 1+3 &0.0060(5)+0.022(2) \\
  &    & &2 &0.200(8) \\
  5357(4)   &  &  &  &   \\
  5386.8(15) & $(3^{-},4,5,6^{-})$   & *  & 3 &0.0128(7) \\
  5424(4) &  &  &  &     \\
  (5436.2(20))&    & 5435(2)  & 0 &0.0110(6) \\
  5444(4) &  &    &  &   \\
  (5454.2(20))&    & 5453(2)  & 0 &0.0034(5) \\
  5485(4) &  &    &  &   \\
  5540.8(11) & $3^{-}$     &  &  & \\
  5577.5(7)& $(2^{-},3)$  & 5577(2) & 0 &0.081(1) \\
  5606(4) &    & 5608(2) & 3&0.334(3)  \\
  5635.7(2) & $(1,2^{+})$  & *  & 1+3 &0.0081(7)+0.054(2) \\
  & &   & 2 &0.247(4) \\
  5672.9(10) & $(1,2^{+})$   & *  & 0&0.0118(8)  \\
  5705(5) &  &  &  &     \\
  5763.2(10) & $(1^{+},2^{+})$   &    &  & \\
  5785.5(10)& $(2,3,4^{+})$   & 5784(2)   &1  &0.0010(1) \\
  5805.9(10) & $(2^{+})$ &   *  & 1+3 &0.0028(6)+0.016(1) \\
  5852.8(3) & $(3^{-})$   & *  & 1+3 &0.012(1)+0.039(3) \\
  5868(1) &  &  &  &     \\
  5897.1(10) & (2) &   *  & 1 &0.0088(7) \\
  \hline
\end{tabular}
\end{minipage}}
\label{table1} 
\raggedright
\footnotesize{\\$^{*}$ observed and used in energy calibration\\
$^{a}$ observed at only one angle}
\end{table}

\begin{table}[t]
\centering
  \caption{Resonance parameters for the $^{33}$S($p,\gamma$)$^{34}$Cl reaction.  The first two columns list adopted $E_{x}$ and $E_{R}$, as determined using a weighted average of present and previous work (see Table \ref{table1}) and $S_{p}$($^{34}$Cl)=5143.2 keV\cite{AME12}.  The third column gives adopted $J^{\pi}$ values as found through previous and present constraints.  Previously measured ($p,\gamma$) resonance strengths $\omega\gamma$ are listed in the fourth column.  The last two columns list proton partial widths $\Gamma_{p}$ corresponding to the adopted orbital angular momentum transfer $L$, as determined in the present work. }
\vspace{4 mm}
\resizebox{0.7\textwidth}{!}{\begin{minipage}{\textwidth}
\begin{tabular}{|lll|l|ll|}
  \hline
  \multicolumn{3}{|c|}{Adopted}&Previous work&\multicolumn{2}{|c|}{Present work} \\
  E$_{x}$ (keV) &E$_{R}$ (keV)&J$^{\pi}$ & $\omega \gamma$ (meV)\cite{Fal13,Waa83} & L& $\Gamma_{p} $(meV) \\
  \hline
  5154(4)&11(4)&&&&\\
  5171.6(3) &28.4(3)&4&    &3 & $1.76\times10^{-33}$   \\
   &&&    &2 & $3.50\times10^{-31}$   \\
  5233(4) &90(4) && &   &     \\
  5246(2) & 103(2)&$(2-5)^{-}$&   &3  & $4.27\times10^{-13}$   \\
  5263(4) &120(4) &&   & &      \\
  5278(7) &135(7) &$(5-7)^{+}$&   &  &     \\
  (5292) &(149) &&   &  &    \\
  5314.93(17) &171.73(18)& $7^{+}$ &  &&       \\
  5326(4) &183(4) &&  $<1.4 \times 10^{-3}$ & &      \\
  5338(2)&195(2)  &$(0-5)$&  &1+3 & $2.72\times10^{-5}$    \\
  &  &&  &2 & $2.82\times10^{-5}$    \\
  5357(4) &214(4) & & $<1.0 \times 10^{-2}$  &  &     \\
  5386.8(15) &243.6(15)& $(3-5)^{-}$ & $<2.2\times 10^{-2}$ & 3 & $2.14\times10^{-6}$  \\
  5424(4) &281(4) &&  $<1.0 \times 10^{-2}$ &  &    \\
  5435.4(14)&292.2(14) & $(1,2)^{+}$&  $8.2_{-2.1}^{+1.8} \times 10^{-2}$ &0  & 0.109   \\
  5444(4) &301(4) &&   &  &    \\
  5453.4(14) &310.2(14) & $(1,2)^{+}$& $2.8_{-0.8}^{+1} \times 10^{-2}$ & 0 & $7.57\times10^{-2}$   \\
  5485(4) &342(4) & & $<1.4 \times 10^{-2}$ &  &     \\
  5540.8(11) &397.6(11) &$3^{-}$ & $<4.0 \times 10^{-3}$ & &      \\
  5577.5(7) &434.3(7)&$(1,2)^{+}$  &50(8) &0  &146     \\
  5607.4(18) &464.2(18) & $(2-5)^{-}$&  & 3 & 0.477   \\
  5635.7(2) &492.5(2)&$(1,2^{+})$ & 65(10) & 1+3 &19.5   \\
  && & & 2 &23.1   \\
  5672.9(10) &529.7(10)&$(1,2)^{+}$& 88(38)   & 0 & 196   \\
  5705(5) & 562(5)&& &    &   \\
  5763.2(10) &620.0(10)&$(1,2)^{+}$& 8(4)     &&   \\
  5785.3(9)&642.1(9)&$(2,3)^{-}$& 50(25)   &1 &36.6    \\
  5805.9(10) &662.7(10)&$(0-5)^{-}$ &50(25)  & 1+3 &143     \\
  5852.8(3)&709.6(3) &$(3^{-})$ & 63(25) & 1+3&$1.15\times10^{3}$     \\
  5868(1) &725(1)&    &  &&     \\
  5897.1(10)& 753.9(10)&$(2^{-})$& 63(25)   & 1 & $1.44\times10^{3}$   \\
  \hline
\end{tabular}
\end{minipage}}
\label{table2}  
\end{table}

\begin{table}[]
  \caption{Mean composition of the ejecta (for species in the range S -- Ca, as mass fractions) from 1.35 M$_{\odot}$ oxygen-neon white dwarf nova models adopting the low and high $^{33}$S($p,\gamma$)$^{34}$Cl rates from the present work (see Fig. \ref{fig_rate}).  }
  \begin{center}
\footnotesize{
\begin{tabular}{ |l|l|l|l| }
  \hline
  Nuclide & Low rate & High rate \\
  \hline
$^{32}$S&$5.3\times10^{-2}$&$5.3\times10^{-2}$ \\
$^{33}$S&$4.8\times10^{-4}$&$4.0\times10^{-4}$ \\
$^{34}$S&$4.5\times10^{-4}$&$4.9\times10^{-4}$ \\
$^{35}$Cl&$5.5\times10^{-4}$&$5.8\times10^{-4}$ \\
$^{37}$Cl&$2.0\times10^{-4}$&$2.1\times10^{-4}$ \\
$^{36}$Ar&$7.5\times10^{-5}$&$8.0\times10^{-5}$ \\
$^{38}$Ar&$2.9\times10^{-5}$&$3.0\times10^{-5}$ \\
$^{39}$K&$6.1\times10^{-6}$&$6.2\times10^{-6}$ \\
$^{40}$Ca&$3.1\times10^{-5}$&$3.1\times10^{-5}$ \\
  \hline
\end{tabular}
}
\label{table3}  
  \end{center}
\end{table}

\end{document}